# Sub millimetre flexible fibre probe for background and fluorescence free Raman spectroscopy

Stephanos Yerolatsitis*,1 | András Kufcsák2 | Katjana Ehrlich2 | Harry A. C. Wood1 | Susan Fernandes2 | Tom Quinn2 | Vikki Young2 | Irene Young2 | Katie Hamilton2 | Ahsan R. Akram2 | Robert R. Thomson3 | Keith Finlayson2 | Kevin Dhaliwal2 | James M. Stone1

1 Department of Physics, University of Bath, Claverton Down, Bath, BA2 7AY, UK

2 Translational Healthcare Technologies Team, Centre for Inflammation Research, Queen's Medical Research Institute, University of Edinburgh, Edinburgh, EH16 4TJ, UK

3 Scottish Universities Physics Alliance (SUPA), Inst. of Photonics and Quantum Sciences, Heriot-Watt University, Edinburgh, EH14 4AS UK

* Correspondence
Stephanos Yerolatsitis, Department of Physics, University of Bath, Claverton Down, Bath, BA2 7AY, UK

Email: s.yerolatsitis@bath.ac.uk

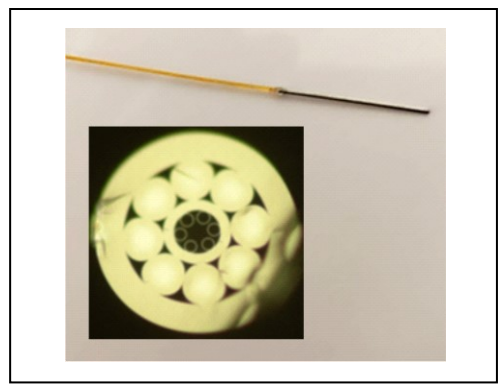

Using the shifted-excitation Raman difference spectroscopy technique and an optical fibre featuring a negative curvature excitation core and a coaxial ring of high numerical aperture collection cores, we have developed a portable, background and fluorescence free, endoscopic Raman probe. The probe consists of a single fibre with a diameter of less than 0.25 mm packaged in a sub-millimetre tubing, making it compatible with standard bronchoscopes. The Raman excitation light in the fibre is guided in air and therefore interacts little with silica, enabling an almost background free transmission of the excitation light. In addition, we used the shifted-excitation Raman difference spectroscopy technique and a tunable 785 nm laser to separate the fluorescence and the Raman spectrum from highly fluorescent samples, demonstrating the suitability of the probe for biomedical applications. Using this probe we also acquired fluorescence free human lung tissue data.

## 1 | INTRODUCTION

Raman spectroscopy is a label-free, non-invasive and non-destructive method for examining unknown samples and obtaining useful information regarding their biochemical composition. The use of Raman spectroscopy as a clinical *in-vivo* tool has not yet been fully realised due to various technical challenges [1]. One such challenge is the silica background generated from the optical fibres used to remotely examine otherwise unreachable tissue regions [2]. Another significant drawback is the autofluorescence background generated from these tissue regions [3]. When examining biomedical samples, the characterisation relies on identifying small and subtle variations in the Raman spectrum. Raman is an inherently weak process (only 1 in $10^6$ photons), consequently, regardless of its source, background light can overwhelm any Raman signal collected from the sample. Several fibre-optic Raman probes have been evaluated with the aim of improving the collection efficiency at the distal end and the separation or elimination of the silica background. Most of these probes consist of either separate fibre cores for excitation and collection or microstructure fibres and bespoke distal optics [4-6]. Nevertheless, this increases the outer diameter, decreases the flexibility of the probe and limits their use to characterise tissue which requires probe miniaturisation for access. To overcome these limitations, we have developed a sub-millimetre flexible fibre-optic Raman probe for endoscopic use by incorporating a single optical fibre featuring a negative curvature excitation core and a coaxial ring of collection cores packaged in sub-millimetre biocompatible polyimide tubing. In negative curvature fibres (NCFs), the light is guided in air and therefore interacts very little ($10^{-4}$ overlap of the guided mode) with the silica structure [7]. We previously demonstrated that such fibres can minimise the generated silica background and enable an almost silica background free collection of the spectrum from the distal sample using a single fibre and no distal optics [2]. While the silica background is minimised in the collected signal through the use of the NCF, fluorescence from the distal sample still introduces a significant challenge in detecting Raman signals from weak Raman scatterers, such as biological tissues. To suppress the sample fluorescence various solutions have been proposed [8]. Here, the shifted-excitation Raman difference spectroscopy (SERDS) technique has been used to reconstruct the weak Raman signals obtained from highly fluorescent samples such as sesame oil and human lung tissue [3, 9-11]. SERDS utilises the fact that the fluorescence background is unaffected by a small shift in the wavelength of the incident laser light. In contrast, the Raman spectrum consists of peaks which have a characteristic shift with respect to the wavelength of the light source. Therefore, a shift in the wavelength of the incident light causes a shift in the entire Raman spectrum. Two spectra (containing fluorescence background and the Raman signal from the sample) are recorded with slightly different wavelengths of the incident laser light. The two measurements are subtracted from each other so that the fluorescence background is removed. The Raman spectrum is retrieved from the difference spectrum, by finding repeated patterns belonging to the two, slightly shifted Raman spectra. This is achieved by a mathematical method based on recurrence relation [11]. It was previously shown that SERDS can eliminate fluorescence more effectively compared to many numerical and baseline fitting methods [9].

For our experiments, we selected a 785 nm laser diode to minimise the generation of fluorescence interference. A longer wavelength excitation laser would minimise the fluorescence interference even further. However, using a longer wavelength excitation laser would have required the use of an InGaAs spectrometer and it would have negatively affected our signal collection efficiency given their higher noise to signal ratio and lower quantum efficiency [7, 10]. By thermally tuning our excitation wavelength, we collected the Raman spectrum (including the fluorescence background) of a highly fluorescent sample (sesame oil) for a range of excitation wavelengths. Using a reconstruction algorithm to recover a background-free spectrum, we examined the effect of the wavelength separation between the two excitation wavelengths and determined the limitations of our reconstruction method. Finally, we tested our probe on excess human lung tissue and successfully reconstructed the expected tissue Raman spectrum.

## 2 | NEGATIVE CURVATURE ANTI-RESONANT FIBRE

The hollow-core NCF was fabricated using the stack and draw technique [12] and the cross-section of the fibre is shown in Figure 1.

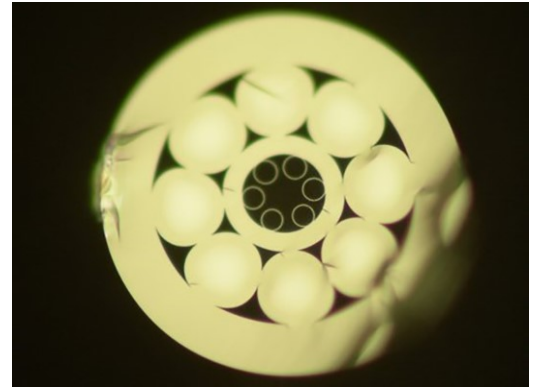

**FIGURE 1** Optical micrograph of the NCF, illuminated by reflected light. The hollow core is bounded by a ring of six capillaries. Surrounding these is a ring of eight solid multimode cores. The outer diameter of the fibre is 210 μm and the core diameter is 20 μm.

The inner region consists of a single ring of six silica capillaries around a hollow core. Depending on the wall thickness, the silica capillaries act as Fabry-Perot resonators, confining the light to the core for specific wavelength bands. For our fibre, the wall thickness was chosen such that the fibre guides light at 785 nm. In addition, surrounding the hollow core we introduced a ring of Ge-doped multimode cores (peak NA = 0.3) designed to collect the Raman signal from the sample at the distal end and therefore to increase the collection efficiency of our probe. It was shown in [2], that by introducing the Ge-doped cores, we achieved a threefold increase in the collection efficiency of our fibre.

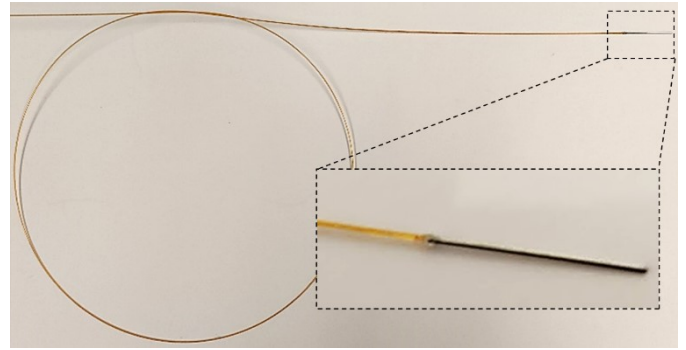

**FIGURE 2** Distal end of our submillimetre probe. The fibre is enclosed inside polyimide tubing. The outer diameter of the tubing is 0.4 mm. The spliced distal tip is protected inside a stainless-steel ferrule

To avoid any capillary action from liquid going through the hollow core of the fibre, we spliced the NCF with a multimode fibre (MMF) with a core diameter matching the collection region of our NCF. We then cleaved the MMF, forming an endcap at the end of our NCF. The fibre was packaged into a 0.4 mm thick biocompatible Polyimide sheath to enclose the fibre (3-metre-long). The distal end with the splice was fixed using a UV curable glue (Loctite 3311) and placed inside a stainless-steel ferrule to increase robustness (Figure 2). By polishing the end face, we managed to reduce the length of the spliced MMF to a few hundred micrometres. Doing so, we minimised the interaction of the light with the silica glass and achieved a high collection efficiency.

## 3 | EXPERIMENTAL SETUP

The probe was connected to our experimental setup as shown in Figure 3, with light coupled to the hollow core using a 785 nm dichroic mirror and a tunable 785 nm distributed Bragg reflector (DBR) laser (DBR785S, Thorlabs).

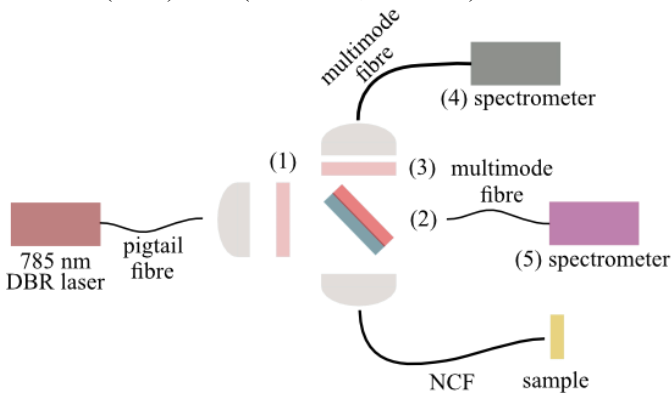

**FIGURE 3** Experimental setup where (1) is a 785 nm bandpass filter (LL01-785-25, Semrock) (2) is a 785 nm dichroic mirror (LPD01-785RS-25, Semrock), (3) is a 791.6 nm longpass filter (LP02-785RU-25, Semrock), (4) is a QE-PRO spectrometer (QePro Raman, Ocean Insight, formerly Ocean Optics) (to collect the backscattered Raman signal from the NCF) and (5) is a spectrometer (Flame-S, Ocean Optics) (to trace the laser's wavelength peak).

The distal end of the fibre was immersed in a highly fluorescent sample (sesame oil). By controlling the temperature of the laser diode using a thermoelectric cooler (TEC) controller, we were able to tune the wavelength of the laser and collect the backscattered Raman signal from our distal sample for a range of excitation wavelengths. The dichroic mirror enables the collection of the backscattered Raman signal from the distal sample. The collected Raman signal was then fibre-fed to an OceanInsight QEPro spectrometer using a 100 μm MMF (peak NA = 0.3). By placing a MMF connected to a second spectrometer (Flame-S, OceanInsight), as shown in Figure 3 (to collect the non-reflected light of the source), we successfully monitored the excitation wavelength of the laser. In our case, the maximum achievable separation is limited by the operational range of our laser diode, approx. 1.5 nm, and the transmission range of our bandpass filter (Figure 3(1)). By tuning the laser between 784.4 nm and 785.9 nm, we investigated the effects of the wavelength separation on the profile of the reconstructed data and discovered the optimum spacing between the two excitation wavelengths. This depended on the wavelength separation and the width of the Raman peaks of the sample and the resolution of our Raman spectrometer (~ 0.37 nm / 6 cm$^{-1}$). The acquisition time was 5s for each collected spectrum and the distal output power was approx. 20 mW.

## 4 | RECONSTRUCTION ALGORITHM

We applied the SERDS technique to purify the Raman spectrum of the sample of interest [9, 11]. As the power of the laser may change when controlling its wavelength and the sample can photobleach with prolonged exposure to light, often the fluorescence intensity is different in the two measured spectra and is not eliminated entirely in the difference spectrum. This can be problematic as the recurrence relation process amplifies the remaining fluorescence background in the difference spectrum. Another undesirable effect of the SERDS technique is that the subtraction of two spectra increases the measurement noise (from the detector and statistical uncertainties) in the difference spectrum, which is then further increased during the reconstruction process. For the former we implemented algorithms suggested by Gebrekidan *et al.* [9], such as z-score normalisation of the measured spectra, and zero-centring the difference spectrum. To account for increased noise, we applied a Savitzky-Golay filter to smooth the zero-centred difference spectrum. After the recurrence relation based reconstruction, baseline correction (using asymmetric least-squares regression on the whole spectrum) and additional smoothing (using a Savitzky-Golay filter) were performed to get the final spectrum. The processing steps are depicted in Figure 4.

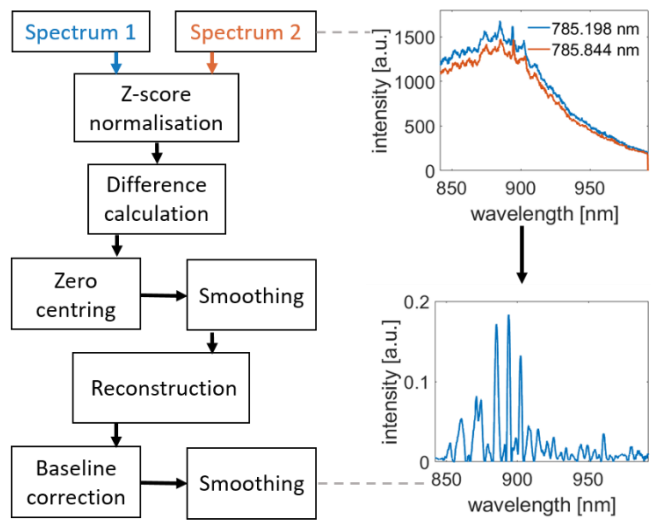

**FIGURE 4** SERDS algorithm processing steps.

The described data processing pipeline can be adapted to the measured data through various parameters of the applied processing steps and filters (e.g. to control the baseline estimation or the amount of smoothing). For direct visual confirmation of the effect of changing these parameters, we implemented an interactive user interface, where the algorithm parameters can be changed (e.g. through slider widgets) and the spectra at different stages of the pipeline is plotted instantly. The ideal settings for the processing steps are selected for each pair of measurements by visual conformation of the quality of the reconstructed spectra with respect to noise, background and spurious signals. The user interface and the entire data pipeline was implemented in Matlab (MathWorks Inc, release 2018b).

## 5 | EXPERIMENTAL RESULTS

To characterise our system, we immersed the fibre probe in sesame oil. We collected the backscattered Raman for a range of excitation wavelengths. At the same time, using a Flame-S spectrometer, we monitored the behaviour of the 785 DBR excitation laser when changing the resistance of the TEC controller. Figure 5 shows the response of the laser for different TEC resistances and therefore temperatures. As we are relying on hopping between one cavity mode to a different one when changing the TEC resistance, the wavelength shift is not linear and smooth. There are temperature ranges where the laser remains on the same cavity mode even when the TEC temperature was changed.

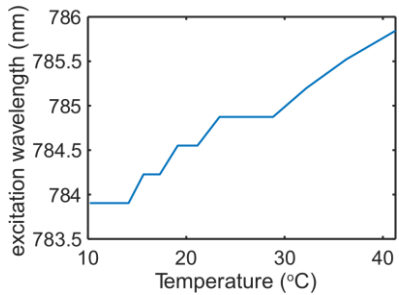

**FIGURE 5** Laser's centre wavelength for different set resistances of the TEC controller.

Taking this into consideration, we chose TEC settings to avoid unstable regions. According to the manufacturer's specifications, we expect a 0.65 nm shift for every 10 °C.
We therefore collected the backscattered Raman signal for various excitation wavelengths (as shown in Figure 6) and reconstructed the spectrum of sesame oil using our algorithm. Sesame oil was chosen as a calibration sample for our system because it exhibits distinct Raman peaks [13, 14] that are stronger than the ones expected from tissue and it has a strong fluorescence background.

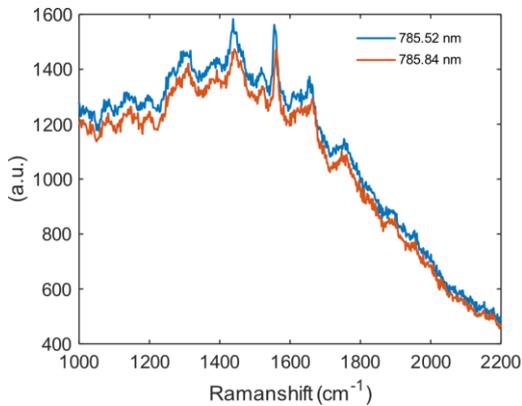

**FIGURE 6** Raw Raman spectra of sesame oil for two excitation wavelengths (0.32 nm separation).

Figure 7 shows the reconstructed spectra of sesame oil using (a) 0.32 nm, (b) 0.65 nm, (c) 0.98 nm, and (d) 1.46 nm separation between the two obtained spectra.

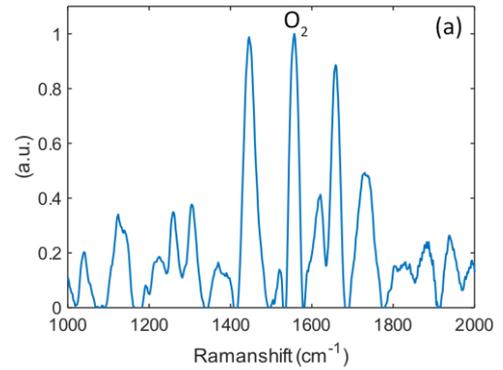
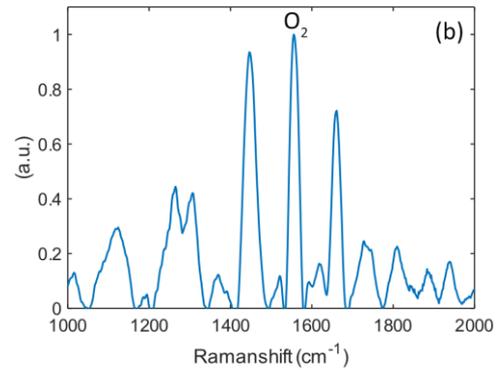
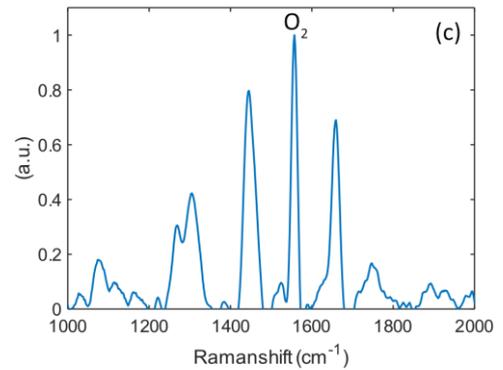
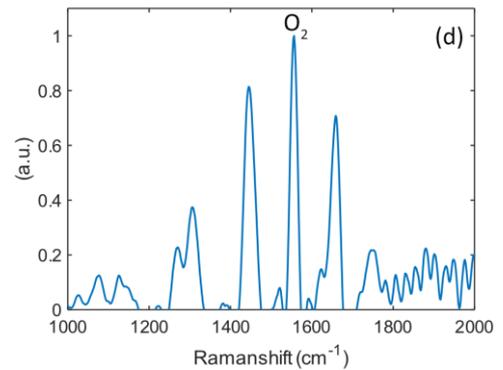

**FIGURE 7** Reconstructed spectra of sesame oil using (a) 0.32 nm, (b) 0.65 nm, (c) 0.98 nm and (d) 1.46 nm separation between the two obtained spectra (5s acquisition time).

Raman peaks are clearly visible at 1268, 1302, 1442, 1660 and 1750 cm$^{-1}$ associated with sesame oil [13, 14], whereas the Raman peak at 1550 cm$^{-1}$ relates to the molecular oxygen from the air inside the hollow core. The oxygen peak was visible in all acquired raw spectra and it was also used as a reference to confirm the wavelength tuning of our excitation laser. In Figure 7, the peaks at 1442 cm$^{-1}$ and 1660 cm$^{-1}$ are visible in all four reconstructed spectra whereas the weaker peaks at 1268 cm$^{-1}$, 1302 cm$^{-1}$ and 1750 cm$^{-1}$ become more visible as the separation between the two excitation wavelengths increases. This relates to the minimum separation that our system can resolve and correlates with the spectral resolution of our spectrometer (0.37 nm). For our system, the optimal excitation wavelength separation is 1 nm (Figure 7c) as at that separation the reconstructed spectrum matches the published spectrum of sesame oil. At longer wavelength separations (i.e. 1.5 nm), there is an increased noise (Figure 7d) due to laser instabilities and the reconstruction method that we used. Using sesame oil, we have demonstrated that the NCF along with the SERDS technique can be used to remove the fluorescence background from a highly fluorescent sample. In addition, the Raman peaks of the reconstructed spectrum match the published spectrum of sesame oil [13, 14].

To demonstrate the capability of our system to recover the intrinsic Raman spectra of human tissue, we performed tests on excess human lung tissue (NHS Lothian Bioresource, REC No. 15/ES/0094). The fresh *ex vivo* tissue samples were obtained from patients diagnosed with suspected or confirmed lung cancer undergoing thoracic resection surgery. We acquired peripheral lung tissue > 5 cm away from the tumour. As shown in Figure 8, we generated a background-free Raman spectrum. The Raman peaks accurately describe proteins and lipids from the distal lung [15].

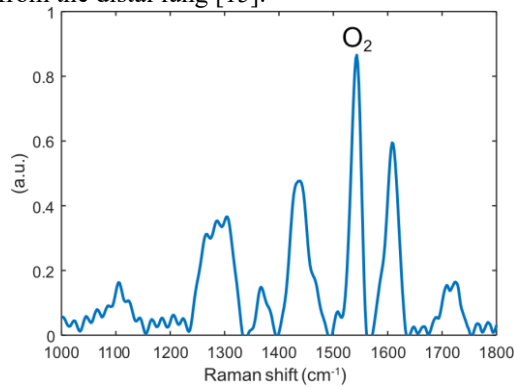

**FIGURE 8** Reconstructed Raman spectrum of human lung tissue (5s acquisition time, 1nm wavelength separation).

## 6 | CONCLUSION

We have developed a background-free Raman probe for endoscopic use by incorporating a single optical fibre featuring a negative curvature excitation core and a coaxial ring of collection cores packaged in sub-millimetre biocompatible polyimide tubing. The probe can be used to collect the Raman signal from distal samples. Using our probe along with the SERDS technique, we have successfully reconstructed the Raman spectrum from a highly fluorescent sample (sesame oil) and from excess human lung tissue. In each case, we were able to identify the Raman peaks related to the sample being studied. In addition, we have observed the effects of the wavelength separation between the excitation wavelengths on the reconstructed spectrum of sesame oil. Although the 5s acquisition time might still be cumbersome for interventional diagnostic procedures, we believe that with further development of the probe and the system to increase the collection efficiency, we can further decrease the required acquisition time. We believe that the development of this system and this miniature fibre-optic Raman probe is a significant step forward for the use of Raman spectroscopy for *in vivo* biomedical applications.

**ACKNOWLEDGMENTS**

The authors thank the BioResource for access to tissue (NHS Lothian BioResource, Scotland Research Ethics Service, reference 15/ES/0094). JS, SY and HW are supported through an EPSRC fellowship (EP/S001123/1). SF is supported through an MRC fellowship (MR/R017794/1). AA is supported by a CRUK Clinician Scientist Fellowship (A24867).

**CONFLICT OF INTEREST**

The authors declare no financial or commercial conflict of interest.

**DATA AVAILABILITY STATEMENT**

Data underlying the results presented in the paper are available https://doi.org/XXXX.